\documentclass{article}

\def\bea{\begin{eqnarray}}
\def\eea{\end{eqnarray}}

\def\pslash{\slash \negthinspace \negthinspace \negthinspace \negthinspace  p}

\newcommand{\reef}[1]{(\ref{#1})}
\newcommand{\rom}[1]{\mathrm{#1}}

\usepackage{amsfonts}
\usepackage{amssymb}
\usepackage{epsfig}

\newcommand{\beqn}{\begin{eqnarray}}
\newcommand{\eeqn}{\end{eqnarray}}

\newcommand{\cn}{{\cal N}}

\newcommand{\be}{\begin{equation}}
\newcommand{\ee}{\end{equation}}

\def\be{\begin{equation}}
\def\ee{\end{equation}}
\def\bea{\begin{eqnarray}}
\def\eea{\end{eqnarray}}
\def\ba{\begin{array}}
\def\ea{\end{array}}
\def\bd{\begin{displaymath}}
\def\ed{\end{displaymath}}

\def\ra{\rangle}
\def\la{\langle}

\def\unit{1 \hskip-.3em \raise2pt\hbox{$ \scriptstyle |$ } }
\def\a{\alpha}

\def\d{\delta}
\def\m{\mu}

\def\bop#1{\setbox0=\hbox{$#1M$}\mkern1.5mu
        \vbox{\hrule height0pt depth.04\ht0
        \hbox{\vrule width.04\ht0 height.9\ht0 \kern.9\ht0
        \vrule width.04\ht0}\hrule height.04\ht0}\mkern1.5mu}
\def\>{\rangle} %right angle
\def\<{\langle} %left angle
\def\Dsl{D \hskip-.6em \raise1pt\hbox{$ / $ } }
\def\to{\rightarrow}

\def\+{\oplus}

\def\lab{\label}
\def\sq2{\sqrt{2}}

\def\ba{{\bar{\a}}}

\def\bd{\bar{\d}}

\newcommand{\lra}{\leftrightarrow}

\begin{document}
%%%%%%%%%%%%%%%%%%%%%%%%%%%%%%%%%%%%%%%%%%%%%

%\thispagestyle{empty}

\begin{titlepage}

\begin{flushright}
MIT-CTP-3880 \\
{\tt hep-th/yymmnnn}\\[3mm]
\today
\end{flushright}
\vspace{1cm}

\begin{center}
{\Large\bf Note on graviton MHV amplitudes}
\vspace{1cm}

{\bf Henriette Elvang${}^{b}$ and
Daniel Z. Freedman${}^{a\,b\,c}$} \\
\vspace{0.7cm}
 { ${}^{a}${\it Department of Mathematics}\\
${}^{b}${\it Center for Theoretical Physics}\\
 {\it Massachusetts Institute of Technology}\\
{\it Cambridge MA 02139, USA} }\\
${}^{c}${\it Currently on leave at CERN, Geneva, Switzerland}\\[5mm]
{\small \tt elvang@lns.mit.edu, dzf@math.mit.edu}

\end{center}
\vskip .3truecm

\begin{abstract}
Two new formulas which express $n$-graviton MHV tree amplitudes in
terms of sums of squares of $n$-gluon amplitudes are discussed. The
first formula is derived from recursion relations. The second formula,
simpler because it involves fewer permutations, is obtained from the
variant of the Berends, Giele, Kuijf formula given in
Arxiv:0707.1035. 
\end{abstract}

\end{titlepage}

%%%%%%%%%%%%%%%%%%%%%%%%%%%%%%%%%%%%%%%%%%%%%%%%%%%

\setcounter{equation}{0}

\section{Introduction}

Spinor-helicity methods have been used in work on gauge theories for
many years. Spinor expressions for S-matrix elements are usually much
simpler than the sum of contributing Feynman diagrams as in the
strikingly simple Parke-Taylor \cite{pt} formula for color ordered  
maximal helicity violating (MHV) gluon amplitudes in tree
approximation: 
\bea\lab{ptf}
  \lab{glueMHV}
  A_n(1^-,2^-,3^+, \dots, n^+) 
  = \frac{\< 1\, 2 \>^4}{ \<1 2 \> \< 2 3 \> \cdots \<n 1 \>} \,.
\eea
The bracket  $\<j\,k\>=-\<k\,j\>$ is the invariant product of positive
helicity spinor 
solutions of the massless Dirac equation for particles of 4-momentum 
$p_j^\m$ and $p_k^\m$.  Much information about the formalism can be
found in reviews such as \cite{dixon, berndixon,shredder}. 
The subject was reinvigorated by the use of twistor ideas
\cite{witten} 
which led to recursion relations \cite{bcfw} for tree amplitudes in
which  
the spinors are treated as complex variables. Feynman diagram
computations can be replaced by the algebraic process of solving the
recursion relations. 

Recursion relations have also been derived for  
tree approximation graviton amplitudes \cite{qmc,cs}, and these are an
important ingredient of this paper.  
MHV amplitudes describe processes
involving two negative and $(n-2)$ positive helicity particles. 
It is well known that these are simpler in both gauge theory and
gravity than non-MHV amplitudes which have more than two negative helicity
particles.
Our primary concern is the set of MHV graviton amplitudes 
$M_n(1^-,2^-,3^+,\dots , n^+)$. 

Our interest in this subject was motivated by recent papers in which
the 3-loop graviton 4-point function was calculated in $\cn=8$
supergravity and shown to be ultraviolet finite \cite{bdr1,bdr2}. The
structures found in the calculation (and in earlier work cited in
these papers)  led the authors to speculate that the S-matrix of
$\cn=8$ supergravity is ultraviolet finite to all orders of
perturbation theory. In the computational approach used in this
program loop amplitudes are 
constructed from tree amplitudes by studying unitarity cuts.  Thus tree approximation amplitudes are a
basic ingredient of higher loop calculations and simplified
expressions for tree amplitudes can be useful. 

The well known KLT relations \cite{klt} express graviton tree
amplitudes $M_n$ in terms of products $A_n\,A_n'$ of gluon amplitudes
in which the momenta in $A_n'$ are a permutation
of those of $A_n$.  The KLT relations for $n=4$ and $n=5$ external
lines are 
\bea \lab{klt4}
M_4(1,2,3,4) &=& -s_{12}\, A_4(1,2,3,4)\, A_4(1,2,4,3) \, ,\\
M_5(1,2,3,4,5) &=& s_{23} \, s_{45} \, A_5(1,2,3,4,5) \, A_5(1,3,2,5,4) 
   + ( 3 \lra 4 ) \,.
\lab{klt5}
\eea
The formulas are more complicated for general $n$. (See Appendix A of \cite{genklt}.)
The KLT relations are valid for all helicity configurations,
and similar formulas relate  amplitudes for any choice of particles      
in supergravity to products of amplitudes  in supersymmetric gauge
theory. In particular tree amplitudes in  $\cn=8$ supergravity  are
related to products of amplitudes for $\cn =4$ gauge theory.

The KLT relations were obtained from string theory. From
the perspective of field theory, however, the relations are very
surprising.   The Lagrangian of Yang-Mills theory, with 3- and 4-point vertices only, appears to be far simpler than the Einstein-Hilbert Lagrangian, which contains complicated
$n$-point two-derivative interactions.
While the 4-point KLT relation has been derived
directly from graviton Feynman rules \cite{sannan}, and field 
redefinitions have been explored \cite{Bern:1999ji,Ananth:2007zy}, no
general field theory derivation has been given.  

The work presented here is a modest step
towards such a derivation and toward the goal of simplified
amplitudes.  We present two formulas for $n$-graviton MHV amplitudes,
each of which expresses  
$M_n(1^-,2^-,3^+,\dots , n^+)$ as a sum of terms containing squares
$A_n(1^-,2^-,i_3^+,\dots , i_n^+)^2$ of gluon amplitudes, where
$i_3,\dots,i_n$ indicates a permutation of the positive helicity
lines. 
The first formula is derived from recursion relations. The complicated
structure of the Lagrangian is thus avoided, but field theoretic
properties such as analyticity and factorization underlie the
recursion relations, and the 
on-shell 3-graviton vertex is required.  The second formula  is
obtained by manipulation of a recently presented version \cite{unexp} 
of the BGK formula \cite{bgk}. 

The  formula derived from recursion relations is (for $n \ge 4$) 
\bea
  M_n(1^-, 2^-, 3^+,\dots ,n^+)
  &=&
  \sum_{\mathcal{P}(i_3,\dots, i_n)}
  s_{1 i_n} \left( \prod_{s=4}^{n-1} \beta_s  \right)
  A_n(1^-, 2^-, i_3^+,\dots ,i_n^+)^2 
  \, , ~~~~~
  \lab{genMn}
\eea
where\footnote{The notation includes spinors $j],\,k]$ which are
  negative helicity solutions of the Dirac equation for null momenta
  $p^\m_j,\,p^\m_k$. They appear     
through   $[j\,k]$ and  $\< j | i |k]$ which are defined  by ($p_i^\m$ is also null) 
\bea 
  \nonumber
  [j\,k]&=& \frac{s_{jk}}{\<k\,j\>} ~=~ -\frac{(p_j +p_k)^2}{\<k\,j\>}
  \, \\
  \nonumber
  \< j |\, i\,|k]&=& \< j |\;\pslash_i\,|k] ~=~\<j\,i\>[i\,k]\, .
\eea
}
\bea
  \lab{beta}
  \beta_s = -\frac{\< i_s\, i_{s+1} \>}{\< 2\, i_{s+1} \>}
    \; \< 2 | \, i_3 + i_4 + \dots + i_{s-1} | i_s ] \, .
\eea
The sum in \reef{genMn} is over all permutations 
$\mathcal{P}(i_3,\dots, i_n)$ of the
external positive helicity labels $\{3,4,\dots , n\}$.
Our new version of the BGK formula is 
\bea
  \nonumber
  M_n(1^-, 2^-, 3^+,\dots ,n^+)
  &=&
  \sum_{\mathcal{P}(i_4,\dots, i_n)}
  \frac{\<1 \, 2 \> \< i_3 \, i_4 \>}{\<1 \, i_3 \> \<2\, i_4 \>} \,
  s_{1 i_n} \left( \prod_{s=4}^{n-1} \beta_s  \right)
  A_n(1^-, 2^-, i_3^+,\dots ,i_n^+)^2 
  \, , ~~~~~~\\
  \lab{bbgk}
\eea
with the same $\beta_s$.  The distinguished line $i_3$ can be any
chosen member of the set $\{3,4,\dots , n\}$, and the sum includes all
permutations of the remaining $n-3$ members. 

The evidence that the formulas above are correct includes:\\[2mm]
i. analytic proof  that \reef{genMn} agrees for all $n$ with the MHV
formula given in  
   \cite{qmc}. \\[1mm]
ii. analytic proof for $n=4,\,5$ that  both  \reef{genMn} and
\reef{bbgk} agree, and also agree with the 
   KLT results \reef{klt4}-\reef{klt5}. \\[1mm]
iii.  numerical work showing that   \reef{genMn} agrees with the
original BGK formula \cite{bgk} for all $n \le 12$.\\[1mm]
iv.  numerical tests of the agreement between  \reef{genMn} and \reef{bbgk} for
     all $n \le 12$ and additional tests that different choices of
     $i_3$ in  \reef{bbgk} do 
      not change the result.

The derivation of \reef{genMn} follows the approach of \cite{qmc} to
recursion relations, but we organize permutations differently and use
gauge theory recursion relations to simplify the work and the result.
This is presented in Sec.~2. In Sec.~3 the passage from the BGK
formula of \cite{unexp} to \reef{bbgk} is outlined. It would be
interesting and useful to extend the treatment of recursion relations
to non-MHV amplitudes, but this is much more difficult.  Our progress
here is limited to a formula for the anti-MHV 5-point function
$M_5(1^-,2^-,3^-,4^+,5^+)$ presented in Sec.~4.

\setcounter{equation}{0}
\section{Derivation of MHV formula \reef{genMn}}

The simple elegant theory underlying recursion relations has been
described clearly in \cite{bcfw,qmc,cs}, so we dispense with the
background and start with the elements we need. Recursion relations
require a shift of either the $|\,j\,]$ or $|\,j\,\>$ spinor of a pair
  of momenta in $n$-point tree amplitudes.  We follow \cite{qmc} and
  use a $[2,1\rangle$-shift, i.e.
\bea
  \lab{shift}
  |\hat{1}\> = |1\> - z |2\> \, , ~~~~
  |\hat{1}] = |1] \, , ~~~~
  |\hat{2}] = |2] + z |1] \, , ~~~~
  |\hat{2}\> = |2\> \, .
\eea
Recursion relations are valid if the analytically continued amplitude vanishes at large $z$,   
 and this property holds for $(--)$ shifts for gluons \cite{bcfw} and for MHV gravitons \cite{bbvc}. 

With this choice, the gluon and graviton MHV  recursion relations
become particularly simple. The gluon recursion relation contains the
single term 
\bea
  \lab{gluonrec}
  A_n(1^-,2^-,3^+,\dots,n^+) 
  &=& A_3(\hat{1}^-,-P_{\hat{1}n}^+,n^+)
  \,\frac{1}{s_{1n}} \,
  A_{n-1}(P_{\hat{1}n}^-,\hat{2}^-, 3^+,\dots,(n-1)^+) \, ,
\eea
since color order must be preserved. 
The graviton recursion relation 
\bea
  \nonumber
  &&M_n(1^-,2^-,3^+,\dots,n^+) 
  \\ 
  && ~~~~~
  = \sum_{\mathcal{P}_c(i_3,\dots ,i_n)} 
  M_3(\hat{1}^-,-P_{\hat{1}i_n}^+,i_n^+)
  \,\frac{1}{s_{1 i_n} } \,
  M_{n-1}(P_{\hat{1}i_n}^-, \hat{2}^- , i_3^+,\dots,i_{n-1}^+)
  ~~~~~\lab{gravrec}
\eea
contains one term for each of the positive helicity lines. (The sum is
  over the \emph{cyclic} permutations of these lines.)

In the recursion relations \reef{gluonrec}-\reef{gravrec} 
each term is evaluated at the value of $z$ that takes the shifted
momentum $P_{\hat{1}k}^\m$ on-shell. 
Hence   
\be 
  0 \,=\, P_{\hat{1}k}^2 \,=\, \<\hat{1}\, k\>[1\, k] 
  \,=\, \big( \<1\, k\>-z \< 2\, k \> \big) \, [1 \, k] \, ,
\ee
determines the value 
\be 
  \lab{zval}
  z =  \frac{\< 1 k \>}{\< 2 k \>} \,.
\ee

The formula \reef{genMn} can be established by an inductive argument
using the fact that $M_3$ and $A_3$ are simply related by
\be  
 \lab{3to3}
 M_3(1^-,-P_{\hat{1}k}^+,j^+)=  A_3(1^-,-P_{\hat{1}k}^+,j^+)^2.
\ee
The basis of induction is established by showing that our formula
reproduces the KLT result for $n=4$. This is done at the end of the
section. 
We assume that \reef{genMn} holds for $M_n$, and then use the
recursion relation for $M_{n+1}$ as follows:
\bea
  \nonumber
  && \hspace{-1cm} 
  M_{n+1}\big(1^-, 2^-, 3^+,\dots ,(n+1)^+\big) \\[3mm]
  \nonumber
  &=&
  \frac{1}{(n-2)!} 
  \sum_{\mathcal{P}(i_3,\dots ,i_{n+1})}
  M_3(\hat{1}^-,-P_{\hat{1}i_{n+1}}^+,i_{n+1}^+) 
  \, \frac{1}{s_{1i_{n+1}}} \,
  M_n(P_{\hat{1}i_{n+1}}^-, \hat{2}^-, i_3^+,\dots ,i_n^+) \, .
  \nonumber
\eea
Bose symmetry of $M_n$ under exchange of any two positive helicity
lines was used to turn the sum over cyclic permutations in
  \reef{gravrec} into a sum over all permutations. The factor
  $1/(n-2)!$ compensates the overcounting.  

The formula \reef{genMn} is now substituted for the $n$-point
graviton amplitude, and \reef{3to3} is used to write $M_3 =
A_3^2$. Then
\bea
   \nonumber
  && \hspace{-1cm} 
  M_{n+1}\big(1^-, 2^-, 3^+,\dots ,(n+1)^+\big) \\[3mm]
  \nonumber
  &=&
  \frac{1}{(n-2)!} \sum_{\mathcal{P}(i_3,\dots ,i_{n+1})}
  A_3(\hat{1}^-,-P_{\hat{1}i_{n+1}}^+,i_{n+1}^+)^2 
  \, \frac{1}{s_{1i_{n+1}}}  \,  \\[1mm]
 \nonumber
  &&\hspace{3.8cm} \times
  \sum_{\mathcal{P}(i_3,\dots ,i_n)}
  s_{i_n\, P_{\hat{1}i_{n+1}}}
  \left( \prod_{s=4}^{n-1} \beta_s  \right)
  A_n(P_{\hat{1}i_{n+1}}^-, \hat{2}^-, i_3^+,\dots ,i_n^+)^2\\[3mm] 
  \nonumber
  &=&
  \sum_{\mathcal{P}(i_3,\dots ,i_{n+1})} 
  A_3(\hat{1}^-,-P_{\hat{1}i_{n+1}}^+,i_{n+1}^+)^2 
  \, \frac{1}{s_{1i_{n+1}}}  \, %\\
  \nonumber
  s_{i_n\, P_{\hat{1}i_{n+1}}}  
  \left( \prod_{s=4}^{n-1} \beta_s  \right)
  A_n(P_{\hat{1}i_{n+1}}^-, \hat{2}^-, i_3^+,\dots ,i_n^+)^2 \\[3mm]
  &=&
  \sum_{\mathcal{P}(i_3,\dots ,i_{n+1})} 
  s_{1i_{n+1}} \,
  s_{i_n\, P_{\hat{1}i_{n+1}}}
  \left( \prod_{s=4}^{n-1} \beta_s  \right) \,
  A_{n+1}\big(1^-, 2^-, i_3^+,\dots ,i_{n+1}^+\big)^2 \, .
  \lab{mnplus1}
\eea
The factor $1/(n-2)!$
cancels because of the redundant inner permutation sum. 
In the last line we use the gauge theory recursion relation
  \reef{gluonrec} to replace the product $A_3 A_n$ by  $s A_{n+1}$.

The final step in the proof is to show that 
$s_{i_n\,  P_{\hat{1}i_{n+1}}} = \beta_{n}$.
Recall that $P_{\hat{1}i_{n+1}}^\m$ is a null vector with $z$
evaluated as in \reef{zval}, i.e.~ 
$z = \la 1 \,  i_{n+1} \ra / \la 2\,  i_{n+1} \ra$.
Then, using that $P_{\hat{1}i_{n+1}}^2=0$, we have 
\bea
  \nonumber
  s_{i_n\,  P_{\hat{1}i_{n+1}}} &=&
  -\Big(p_{i_n} + (p_{\hat{1}}+p_{i_{n+1}} ) \Big)^2 \\[2mm]
  \nonumber
  &=& 
  -2 \, p_{i_n} \cdot p_{\hat{1}} 
  - 2 \, p_{i_{n+1}} \cdot p_{i_n} \\[2mm]
  \nonumber
  &=& - \la \hat{1} \, i_n  \ra [ \hat{1} \, i_n ]
      - \la i_{n+1} \, i_n \ra [ i_{n+1} \, i_n] \\[2mm]
  \nonumber
  &=&
  -\frac{[1 \, i_n]}{\la 2\,  i_{n+1} \ra}
  \Big( 
      \la 1\,  i_n  \ra \la 2 \,  i_{n+1} \ra
    - \la 2\, i_n  \ra  \la 1 \, i_{n+1} \ra 
  \Big) 
  - \la i_{n+1} \, i_n \ra [ i_{n+1} \, i_n] \\[2mm]  
  \nonumber
  &=& 
  \frac{\< i_n \, i_{n+1} \> }{\la 2 \,  i_{n+1} \ra}
  \Big( 
    \la 2 1 \ra [ 1 \, i_n ]
    + \la 2 \, i_{n+1} \ra [ i_{n+1}\, i_n ]
  \Big) \\[2mm]
  \nonumber
  &=& 
    \frac{\< i_n \, i_{n+1} \> }{\la 2  \, i_{n+1} \ra}
  \; \< 2 | \,1 + i_{n+1} | i_n ]\\[2mm]
    \nonumber
  &=& 
  -\frac{\< i_n \, i_{n+1} \> }{\la 2 \,  i_{n+1} \ra}
  \< 2 | \, i_3 + i_4 + \dots +  i_{n-1} | i_n ]\\[2mm]
  &=& 
  \beta_{n} \, .
  \lab{evals}
\eea
We used the Schouten identity in the 5th line and momentum
conservation in the last step.  
This establishes 
\reef{genMn} for $M_{n+1}$, and the inductive proof is complete.

Let's examine the cases $n=4, 5$ of \reef{genMn} in more detail. 
For $n=4$, the product in \reef{genMn} is over the empty set and is
set equal to 1. One then finds
\bea
  \lab{our4}
  M_4(1^-, 2^-, 3^+, 4^+)
  &=&
  s_{14}\, A_4(1^-, 2^-, 3^+, 4^+)^2 + ( 3 \lra 4) \, .
\eea
Using the explicit form of gluon tree amplitudes \reef{glueMHV} one
can show (using momentum conservation) that $A_4(1^-, 2^-, 3^+, 4^+)$
differs from $A_4(1^-, 2^-, 4^+, 3^+)$ differ by a simple factor of
$s_{13}/s_{14}$, and hence \reef{our4} gives
\bea
  M_4(1^-, 2^-, 3^+, 4^+)
  &=&
  \Big(  s_{14}\, \frac{s_{13}}{s_{14}} 
        + s_{13}\, \frac{s_{14}}{s_{13}}\Big)  \,
  A_4(1^-, 2^-, 3^+, 4^+) \, A_4(1^-, 2^-, 4^+, 3^+) \, .
\eea
The KLT result \reef{klt4} then follows from $s_{12} + s_{13} + s_{14}
= 0$. 

{}For $n=5$,
\bea
  \prod_{s=4}^{n-1} \beta_s &=& \beta_4 
          ~=~ - \frac{\< i_4\, i_5 \>}{\< 2\, i_5 \>}
               \; \< 2 | \, i_3 | i_4 ] 
          ~=~ - \frac{\< i_4\, i_5 \>}{\< 2\, i_5 \>}  
               \; \< 2 \, i_3 \> [i_3 \, i_4 ]  \, .
\eea
Using this one can show analytically that \reef{genMn} reproduces the
KLT result \reef{klt5}.

%%%%%%%%%%%%%%%%%

\subsection{Connection to the graviton MHV formula of \cite{qmc}}

The result of \cite{qmc} for MHV graviton amplitudes is 
\bea
  \nonumber
  && \hspace{-1cm} M_n(1^-, 2^-, 3^+,\dots ,n^+) \\[2mm]
  \nonumber
  &=&
  (-1)^{n+1} \sum_{\mathcal{P}(i_3,\dots, i_n)}
  \frac{\<1\, 2 \>^6\,  [1 \, i_n]}{\<1 \, i_n \>} \,
  \frac{1}{2} \,
  \frac{[i_3 \, i_4]}
  {\< 2 \, i_3 \> \< 2\,  i_4 \> \< i_3 \, i_4 \>
    \< i_3\, i_5 \> \< i_4\, i_5 \> }
  \left( \prod_{s=5}^{n-1} 
         \frac{\< 2 | \, i_3 + \dots +  i_{s-1} | i_s ]}
              {\< 2\, i_{s+1} \>\< i_s\, i_{s+1} \>}\right) \, . \\
  \lab{QMC}
\eea

It is not difficult to obtain \reef{QMC} from \reef{genMn}. We write 
the gauge theory MHV amplitude as
\bea
 A_n\big(1^-, 2^-, i_3^+,\dots ,i_n^+\big)
 = \frac{\<1\,2\>^3}{\< 2\, i_3 \> \< i_3 \, i_4 \> 
     \Big(\prod_{s=4}^{n-1} \< i_s \, i_{s+1} \> \Big) 
     \< i_n \, 1 \> } \, .
\eea
Substitute this into the MHV relation \reef{genMn} and use 
$s_{1 i_n} = -  \<1 \, i_n \> [1 \, i_n]$.  Then
\bea
  \nonumber
  && \hspace{-1cm} M_n(1^-, 2^-, 3^+,\dots ,n^+) \\
  \nonumber
  &=&
  (-1)^{n}\sum_{\mathcal{P}(i_3,\dots, i_n)}
  s_{1 i_n} 
  \left( \prod_{s=4}^{n-1} \frac{\< i_s\, i_{s+1} \>}{\< 2\, i_{s+1} \>}
    \; \< 2 | \, i_3 + \dots +  i_{s-1} | i_s ]  \right)
  A_n(1^-, 2^-, i_3^+,\dots ,i_n^+)^2  \\[2mm]
  &=&
  (-1)^{n+1}
  \sum_{\mathcal{P}(i_3,\dots, i_n)}
  \frac{\<1\, 2 \>^6\,  [1 \, i_n]}
       {\< 2 \, i_3 \>^2 \< i_3 \, i_4 \>^2 \<1 \, i_n \>}
  \frac{\< 2 | \, i_3 | i_4]}
       {\< 2\, i_5 \>\< i_4\, i_5 \>}
  \left( \prod_{s=5}^{n-1} 
         \frac{\< 2 | \, i_3 + \dots +  i_{s-1} | i_s ]}
              {\< 2\, i_{s+1} \>\< i_s\, i_{s+1} \>}\right) \, .
\eea
Using that $\< 2 | \,i_3 | i_4] = \< 2 \, i_3 \> [i_3 \, i_4]$, we
  find  
\bea
  \nonumber
  && \hspace{-1cm} M_n(1^-, 2^-, 3^+,\dots ,n^+) \\
  \nonumber
  &=&
  (-1)^{n+1}
  \sum_{\mathcal{P}(i_3,\dots, i_n)}
  \frac{\<1 \,2 \>^6\,  [1 \, i_n]}{\<1 \, i_n \>} 
  \frac{[i_3 \, i_4]}
       {\< 2 \, i_3 \>  \< 2\, i_5 \> \< i_3 \, i_4 \>^2 \< i_4\, i_5 \>}
  \left( \prod_{s=5}^{n-1} 
         \frac{\< 2 | \, i_3 + \dots +  i_{s-1} | i_s ]}
              {\< 2\, i_{s+1} \>\< i_s\, i_{s+1} \>}\right) \, .\\
  \lab{nearQMC}
\eea
This is not quite the result \reef{QMC}. Note though that under
exchange of $i_3$ and $i_4$, the product $\prod$ is invariant. Since
we are summing over all permutations of the positive helicity lines
$i_k$, we can include explicitly the $i_3 \lra i_4$ permutation and
divide by 2 to compensate for the overcounting. This allows us to
rewrite \reef{nearQMC} as
\bea
  \nonumber
  \frac{[i_3 \, i_4]}
       {\< 2 \, i_3 \>  \< 2\, i_5 \> \< i_3 \, i_4 \>^2 \< i_4\, i_5 \>}
  &\to&
  \frac{1}{2}
  \frac{[i_3 \, i_4]}
  {\< 2\, i_5 \> \< i_3 \, i_4 \>^2}
  \left( \frac{1}{ \< 2 \, i_3 \>\< i_4\, i_5 \>}
        -\frac{1}{ \< 2 \, i_4 \>\< i_3\, i_5 \>} \right) \\[2mm]
  \nonumber
  &=&
  \frac{1}{2}
  \frac{[i_3 \, i_4]}
  {\< 2 \, i_3 \> \< 2\,  i_4 \> \< i_3 \, i_4 \>
    \< i_3\, i_5 \> \< i_4\, i_5 \> } \, ,
\eea
by the Schouten identity. This gives \reef{QMC} exactly.

%%%%%%%%%%%%%%%%%%%%%%%%%%%%%%%%%%%%%%%%%%%%%%%%%%%%%%

\section{BGK as (gauge theory)$^2$}
The authors of \cite{unexp} presented the BGK formula in a simpler
form, which we write here as
\bea
  M_n = - \<a\, b\>^8
  \sum_{\mathcal{P}(i_4,\dots,i_n)}
  \frac{\prod_{s=4}^{n} \<n |\, 2 + i_4 + i_5+\dots + i_{s-1} | \, i_s]}
  {\< 1\,i_n \> \<1\, n \>^2 \<2\, n \>^2 \< 1\,2\> \<2 \, i_4 \>\< i_n \, n \>
   \big(\prod_{s=4}^{n-1} \< i_s \, i_{s+1}\> \< i_s \, n \>\big)} \, .
   \lab{bgk07}
\eea
The external lines are $(1^+,2^+,\dots,a^-,\dots,b^-,\dots,n^+)$  
and the permutation sum $\mathcal{P}(i_4,\dots,i_n)$ is over momentum
labels $\{3,4,\dots,n-1\}$.  

The formula \reef{bbgk} is a simple rewriting of \reef{bgk07}. First
we relabel the external legs to the effect of interchanging $p_2$ and
$p_n$. Then we select the negative helicity lines to be $a=1$ and
$b=2$, and we introduce $i_3 = n$. Finally we rewrite the products in
\reef{bgk07} to explicitly include the $A_n^2$ factor. The result is
the formula \reef{bbgk}. It is clear that by an initial relabeling of
the external lines, the distinguished line $i_3$ could have been any
one of the positive helicity lines. 

The original BGK formula \cite{bgk} can also be rewritten as a sum
over gluon amplitudes squared, but we have chosen to work with
\reef{bgk07} in order to display the form which most closely 
resembles our formula \reef{genMn}.

%%%%%%%%%%%%%%%%%%%%%%%%%%%%%%%%%%%%%%%%%%%%%%%%%%%%%%

\setcounter{equation}{0}
\section{A modest non-MHV result}

Loop amplitudes in gravity and supergravity require more than MHV tree
 amplitudes as input. For example the non-MHV
 amplitude\footnote{Recursion relations were used in \cite{cs} to
 obtain a spinor helicity formula for this amplitude.} 
 $M_6(1^-,2^-,3^-,4^+,5^+,6^+)$ was needed in the 3-loop calculation
 of \cite{bdr1}.  Thus it would be of both practical and intrinsic
 interest to extend the treatment of recursion relations in Sec.~2 to
 non-MHV amplitudes.  Unfortunately the non-MHV sector is more
 complicated for both gluons and gravitons.  Our results to date are
 limited to a 
new expression\footnote{A spinor helicity formula was given earlier in
 \cite{bore}.} for the anti-MHV amplitude $M_5(1^-,2^-,3^-,4^+,5^+)$
 involving a sum 
over  squares of gluon $A_5$'s.  Of course, this amplitude is the
 complex conjugate of  
the   MHV
$M_5(1^+,2^+,3^+,4^-,5^-)$, and this fact provides a check which the
 formula obtained below satisfies. We present our formula with few
 details as an indication of the complications encountered in the
 non-MHV sector. 

The relevant graviton recursion relation, obtained using a  $[2, 1\>$ shift, is
\bea \lab{5rec}
  \nonumber
  M_5(1^-,2^-,3^-,4^+,5^+)
  &=& 
  \Big\{  M_4(\hat{1}^-,3^-,P_{\hat{2}4}^+, 5^+) 
  \, \frac{1}{s_{24}} \,
  M_3(-P_{\hat{2}4}^-, \hat{2}^-, 4^+) 
  + ( 4 \lra 5) \Big\} \\
  &&
  + M_4(\hat{1}^-,P_{\hat{2}3}^-,4^+, 5^+) 
  \, \frac{1}{s_{23}} \,
  M_3(-P_{\hat{2}3}^+, \hat{2}^-, 3^-)
\eea
Since the right side involves  only  3- and 4-point functions we can
insert the results 
\reef{3to3} and \reef{our4}, with conjugation and shifts as appropriate.
The result is a sum of 
terms involving products $(A_4\,A_3)^2$ for various
configurations of momenta. The 
strategy of Sec.~2 suggests that we use gauge theory recursion
relations to replace 
these products with $(A_5)^2$.  However this is tricky because the
recursion relation for one of the needed orderings of external gluons has
two terms\footnote{The minus sign is 
required because of anti-cyclic ordering in the first term.} 
\bea
  \nonumber
  A_5(1^-,3^-,2^-,4^+,5^+)
  &=& - A_4(\hat{1}^-,P_{\hat{2}3}^-,4^+, 5^+)
  \frac{1}{s_{23}}
  A_3(-P_{\hat{2}3}^+, \hat{2}^-, 3^-) \\
  && +
  A_4(\hat{1}^-,3^-, P_{\hat{2}4}^+, 5^+)
  \frac{1}{s_{24}}
  A_3(-P_{\hat{2}4}^-, \hat{2}^-, 4^+) \, .
  \lab{2terms}
\eea
Nevertheless we use \reef{2terms} and the one-term recursion relations which
hold for other orderings to derive the following representation:
\bea
   \nonumber
 M_5(1^-,2^-,3^-,4^+,5^+)
  &=& 
  \bigg\{ 
   s_{24} \, s_{\hat{1}5} 
     \Big[  A_5(1^-,2^-,3^-,4^+,5^+) + A_5(1^-,3^-,2^-,4^+,5^+) \Big]^2 \\
   \nonumber
   &&  \hspace{5mm}
   + s_{24} \,  s_{35} 
     A_5(3^-,1^-,2^-,4^+,5^+)^2 \\
   &&  \hspace{5mm}
   + s_{23} \, s_{\hat{1}5}
     A_5(1^-,2^-,3^-,4^+,5^+)^2
  \bigg\} + (4 \lra 5) \, ,
  \lab{barM5}
\eea
which essentially does express the  graviton $\overline{\rom{MHV}}$ amplitude in
terms of squares of $\overline{\rom{MHV}}$ gluon amplitudes. Readers with good
eyesight will notice that the invariant $s_{\hat{1}5}$ contains a shift to be evaluated at the
appropriate poles,  
\bea
  P_{\hat{2}4}^2 = 0 ~~~~&\to& ~~~~~
  s_{\hat{1}5} =  \frac{\< 35 \> [15][34]}{[14]} \, , \\
  P_{\hat{2}3}^2 = 0 ~~~~&\to& ~~~~~
  s_{\hat{1}5} =  - \frac{\< 45 \> [15][34]}{[13]} \, .
\eea
These results are used in the first and third line of \reef{barM5},
respectively.   

%%%%%%%%%%%%%%%%%%%%%%%%%%%%%%%%%%%%%%%%%%%%%%%%%%%%%%

\setcounter{equation}{0}
\section{Discussion}

The formulas \reef{genMn} and \reef{bbgk} express graviton MHV
amplitudes $M_n$ as sums of gluon MHV amplitudes $A_n$ squared. This
is a first step towards obtaining general-$n$ KLT-like relations from
field theory. We have proven our formula \reef{genMn} by
induction using recursion relations. The fact that the BGK formula can
be written in a very similar way \reef{bbgk} should facilitate an
analytic proof of the BGK formula. 

It was noted in \cite{unexp} that under a $(-,-)$-shift the BGK
formula \reef{bgk07} behaves as $z^{-2}$ for large $z$. Our rewriting
\reef{bbgk} of \reef{bgk07} clearly exhibits this property too, and it
also makes it manifest that, for this type of shift, the large
$z$-behavior of $M_n$ is identical to that of $A_n^2$. On the other
hand, our formula \reef{genMn} has naively a leading $z^{-1}$
fall-off. We have checked numerically up to $n = 11$ that this leading
term vanishes. This is an indication of the redundancy of the $(n-2)$
extra permutations in \reef{genMn} compared with \reef{bbgk}. 

In the proof of \reef{genMn}, we first used the gravity recursion
relations to express $M_n$ in terms of $M_3$ and $M_{n-1}$ and then
the inductive assumption to get from $M_{n-1}$ to (sum of) $A_{n-1}^2$. A very
useful step was then to use that the gauge theory recursion relations
only contained one term, so that one could replace  
$A_3^2\, A_{n-1}^2$ by $s^2 \,A_n^2$. It is not clear that one can
generalize this step to non-MHV, since as we illustrated in Sec.~3,
the gauge recursion relations will contain several terms. Beyond $n=5$
the $(-,-)$ shift does not seem to make the step  
$A_k^2 A_{n-k+2}^2 \to s^2 \,A_n^2$ possible.  

Tree amplitudes play an important role in loop calculations, and our
work is a step towards deriving useful relations of the form $M_n =
\sum A_n^2$ from field theory.  

%%%%%%%%

\section*{Acknowledgements}
We are indebted to Z.~Bern, L.~Dixon and R.~Roiban for guidance and
encouragement. 
We have also benefited from discussions with A.~Brandhuber,
C.~Berger, M.~Green, 
J.~McGreevy, K.~Risager, M.~Rangamani,  B.~ Spence,  G. ~Travaglini,
and P.~Vanhove.  
Considerable progress was made at the Benasque Center for Science.  HE
would like to thank the Niels Bohr Institute for their hospitality.  
HE is supported by a Pappalardo Fellowship in Physics at MIT, and
DZF is supported by NSF grant PHY-0600465 and by TH-division, CERN.
Both authors are supported by the US Department of Energy through cooperative research agreement DE-FG0205ER41360.

%%%%%%%%%%%%%%%%%%%%%%%%%%%%%%%%%%%%%%%%%%%%%%%


\begin{thebibliography}{99}

%\cite{Parke:1986gb}
\bibitem{pt}
  S.~J.~Parke and T.~R.~Taylor,
  ``An Amplitude for $n$ Gluon Scattering,''
  Phys.\ Rev.\ Lett.\  {\bf 56}, 2459 (1986).
  %%CITATION = PRLTA,56,2459;%%

\bibitem{dixon}
L.~Dixon,
"Calculating Scattering Amplitudes Efficiently,"
Theoretical Advanced Study Institute (TASI 95)

\bibitem{berndixon}
Z.~Bern, L.~J.~Dixon and D.~A.~Kosower,
  ``On-Shell Methods in Perturbative QCD,''
  Annals Phys.\  {\bf 322}, 1587 (2007)
  [arXiv:0704.2798 [hep-ph]].

\bibitem{shredder}
M.~Srednicki
"Quantum Field Theory,"
Cambridge University Press, 2007.

\bibitem{witten}
E.~Witten,
  ``Perturbative gauge theory as a string theory in twistor space,''
  Commun.\ Math.\ Phys.\  {\bf 252}, 189 (2004)
  [arXiv:hep-th/0312171].
  %%CITATION = CMPHA,252,189;%%

\bibitem{bcfw}
R.~Britto, F.~Cachazo and B.~Feng,
  ``New recursion relations for tree amplitudes of gluons,''
  Nucl.\ Phys.\  B {\bf 715}, 499 (2005)
  [arXiv:hep-th/0412308].\\
R.~Britto, F.~Cachazo, B.~Feng and E.~Witten,
  ``Direct proof of tree-level recursion relation in Yang-Mills theory,''
  Phys.\ Rev.\ Lett.\  {\bf 94}, 181602 (2005)
  [arXiv:hep-th/0501052].  

\bibitem{qmc}
J.~Bedford, A.~Brandhuber, B.~J.~Spence and G.~Travaglini,
  ``A recursion relation for gravity amplitudes,''
  Nucl.\ Phys.\  B {\bf 721}, 98 (2005)
  [arXiv:hep-th/0502146].

\bibitem{cs}
  F.~Cachazo and P.~Svrcek,
  ``Tree level recursion relations in general relativity,''
  arXiv:hep-th/0502160.

\bibitem{bdr1}
Z.~Bern, L.~J.~Dixon and R.~Roiban,
  ``Is N = 8 supergravity ultraviolet finite?,''
  Phys.\ Lett.\  B {\bf 644}, 265 (2007)
  [arXiv:hep-th/0611086].

\bibitem{bdr2}
Z.~Bern, J.~J.~Carrasco, L.~J.~Dixon, H.~Johansson, D.~A.~Kosower and R.~Roiban,
  ``Three-Loop Superfiniteness of N=8 Supergravity,''
  Phys.\ Rev.\ Lett.\  {\bf 98}, 161303 (2007)
  [arXiv:hep-th/0702112].

\bibitem{klt}
  H.~Kawai, D.~C.~Lewellen and S.~H.~H.~Tye,
  ''A Relation Between Tree Amplitudes Of Closed And Open Strings,''
  Nucl.\ Phys.\  B {\bf 269}, 1 (1986).
  %%CITATION = NUPHA,B269,1;%%

\bibitem{genklt}
  Z.~Bern, L.~J.~Dixon, M.~Perelstein and J.~S.~Rozowsky,
  ``Multi-leg one-loop gravity amplitudes from gauge theory,''
  Nucl.\ Phys.\  B {\bf 546}, 423 (1999)
  [arXiv:hep-th/9811140].
  %%CITATION = NUPHA,B546,423;%%

\bibitem{sannan}
  S.~Sannan,
  ``Gravity As The Limit Of The Type II Superstring Theory,''
  Phys.\ Rev.\  D {\bf 34}, 1749 (1986).
  %%CITATION = PHRVA,D34,1749;%%

\bibitem{Bern:1999ji}
  Z.~Bern and A.~K.~Grant,
  ``Perturbative gravity from {QCD} amplitudes,''
  Phys.\ Lett.\  B {\bf 457}, 23 (1999)
  [arXiv:hep-th/9904026].

\bibitem{Ananth:2007zy}
  S.~Ananth and S.~Theisen,
  ``KLT relations from the Einstein-Hilbert Lagrangian,''
  Phys.\ Lett.\  B {\bf 652}, 128 (2007)
  [arXiv:0706.1778 [hep-th]].
  %%CITATION = PHLTA,B652,128;%%

\bibitem{unexp}
  Z.~Bern, J.~J.~Carrasco, D.~Forde, H.~Ita and H.~Johansson,
  ``Unexpected Cancellations in Gravity Theories,''
  arXiv:0707.1035 [hep-th].
  %%CITATION = ARXIV:0707.1035;%%

\bibitem{bgk}
  F.~A.~Berends, W.~T.~Giele and H.~Kuijf,
  ``On relations between multi - gluon and multigraviton scattering,''
  Phys.\ Lett.\  B {\bf 211}, 91 (1988).
  %%CITATION = PHLTA,B211,91;%%

\bibitem{bbvc}
  P.~Benincasa, C.~Boucher-Veronneau and F.~Cachazo,
  ``Taming tree amplitudes in general relativity,''
  arXiv:hep-th/0702032.
  %%CITATION = HEP-TH/0702032;%%

\bibitem{bore}
  N.~E.~J.~Bjerrum-Bohr, D.~C.~Dunbar, H.~Ita, W.~B.~Perkins and K.~Risager,
  ``MHV-vertices for gravity amplitudes,''
  JHEP {\bf 0601}, 009 (2006)
  [arXiv:hep-th/0509016].

\end{thebibliography}
\end{document}